\documentclass{article}

\usepackage[utf8]{inputenc}

\usepackage{tismir}
\usepackage{amsmath}
\usepackage[hidelinks]{hyperref}
\usepackage{url}
\usepackage{graphicx}
\usepackage{tabularx}
\usepackage{multirow,multicol}
\usepackage{adjustbox}
\usepackage{array}
\usepackage{float}

\usepackage[autostyle, english = american]{csquotes}
\MakeOuterQuote{"}

\graphicspath{{./imgs/}}

\title{Style-based Composer Identification and Attribution of Symbolic Music Scores: a Systematic Survey}
\titleshort{Style-based Composer Identification and Attribution of Symbolic Music Scores}

\author{Federico Simonetta\thanks{GSSI -- Gran Sasso Science Institute, L'Aquila, Italy}}
\authorref{Simonetta,~F.}

\date{}

\type{overview}

\authorshort{Simonetta} 

\begin{document}
\twocolumn[{%
			\maketitleblock
			\begin{abstract}
This paper presents the first comprehensive systematic review of literature on style-based composer identification and authorship attribution in symbolic music scores. Addressing the critical need for improved reliability and reproducibility in this field, the review rigorously analyzes 58 peer-reviewed papers published across various historical periods, with the search adapted to evolving terminology.
The analysis critically assesses prevailing repertoires, computational approaches, and evaluation methodologies, highlighting significant challenges. It reveals that a substantial portion of existing research suffers from inadequate validation protocols and an over-reliance on simple accuracy metrics for often imbalanced datasets, which can undermine the credibility of attribution claims. The crucial role of robust metrics like Balanced Accuracy and rigorous cross-validation in ensuring trustworthy results is emphasized. The survey also details diverse feature representations and the evolution of machine learning models employed.
Notable real-world authorship attribution cases, such as those involving works attributed to Bach, Josquin Desprez, and Lennon-McCartney, are specifically discussed, illustrating the opportunities and pitfalls of applying computational techniques to resolve disputed musical provenance. Based on these insights, a set of actionable guidelines for future research are proposed. These recommendations are designed to significantly enhance the reliability, reproducibility, and musicological validity of composer identification and authorship attribution studies, fostering more robust and interpretable computational stylistic analysis.
			\end{abstract}
		}]


\section{Introduction}
\label{sec:introduction}

The ILLIAC system by Hiller and Isaacson (1955-1956) was one of the first applications of information technology in music \citep{hillerjr.1958musical}. Around the same time, Youngblood published the first statistical analysis of compositional styles \citep{youngblood1958style}. The field of computational stylistic analysis has since evolved, encompassing diverse literatures and overlapping with tasks like genre recognition and algorithmic composition. Youngblood's work highlights the longstanding significance of this topic in music computing, with important applications in musicology and database labeling \citep{vannuss2017melody}.

In this work, computational methods applied to symbolic scores are considered in the attempt to systematically analyze the literature about composer identification.

The task of identifying the composer of a musical composition based on its symbolic score has been explored from various research perspectives. Computer scientists and statisticians refer to it as "composer identification" or "composer classification," while musicologists and historians call it "authorship attribution."
Despite differing aims, the typical scenario remains the same: given a set of music scores, develop an algorithm to predict the composer of a new score.

The synthesis of the surveyed literature reveals several distinct research perspectives from which the composer identification task is approached.
One reason why scientists approach the composer identification task is to assess the ability of a model to capture the stylistic characteristics of a composer.
In this case, composer identification is often associated with genre recognition and style or epoch classification.
Another application is the labeling of large databases, often mined from the web, that present a large number of music documents but with noisy or erroneous metadata.
In this latter case, the task of composer identification is associated to generic document tagging.
In other works, the task is presented as a benchmark for comparing different feature extractors or architecture.
When one of these scenario occurs, the main focus is usually on the computational methods, with the final task hidden behind generic nomenclature such as ``music tagging'' or ``music classification''.

The aim of authorship attribution is instead to infer the author of a piece of music, given a set of possible authors. This task is often associated with the re-attribution of works whose authorship is uncertain or disputed.
A more in-depth discussion of what is authorship attribution is in Section~\ref{sec:authorship-definition}. From a machine learning perspective, the authorship attribution problem may be regarded as a particular case of the composer identification task.

In this work, the existing literature about composer identification from symbolic music scores is critically reviewed, providing details of the methodology and criteria adopted during the analyses.
Specifically, three main research questions motivated the study:
\begin{enumerate}
	\item \textit{What computational approaches have been developed and applied to composer identification of symbolic music scores?}
	\item \textit{To what extent can music composer identification methods based on compositional style be considered reliable for authorship attribution?}
	\item \textit{How can the research community improve the reliability and reproducibility of music authorship attribution studies?}
\end{enumerate}

Many studies treat "composer identification" as a machine‐learning benchmark -- exploiting gross stylistic differences (e.g., Baroque vs. Romantic) to compare feature extractors or architectures -- without any claim to musicological authorship attribution.
By contrast, the primary focus is on those works whose aim is to support musicological re-attribution of disputed or anonymous pieces, relying on subtle, individual stylistic traits rather than broad period or genre cues.
Nonetheless, the two tasks are closely related, and the distinction is not always clear-cut. While I have attempted to navigate this distinction throughout the analysis, it is often challenging to definitively categorize studies; many works, even when framing composer identification primarily as a benchmarking task, present findings or interpretations that could be perceived as having musicological relevance or broader implications for understanding style, making a rigid separation problematic. Moreover, "composer identification" models may still be used for "authorship attribution" and their study is valid for having a better understanding of the literature and try to answer the research questions.
My critique throughout this survey, therefore, evaluates the reliability and applicability of methods primarily through this lens of musicological authorship attribution.

While a systematic approach that enforces an objective view on the existing literature was adopted, the heterogeneity of the applications and of the nomenclature used in literature made the effort of collecting existing works particularly difficult. Nonetheless, the first comprehensive analysis of the field is provided, achieving a greater understanding of the possibilities that computational techniques open for the re-attribution of existing works.

In Section~\ref{sec:authorship-definition}, the concept of authorship attribution and its similarities and differences with the more general task of composer identification are discussed. In Section~\ref{sec:literature}, the systematic collection of papers is described. In Section~\ref{sec:evaluation}, the evaluation methodologies adopted in the literature are reviewed and a classification among reliable and unreliable protocols is defined. In Sections~\ref{sec:repertoire},~\ref{sec:representations}, and~\ref{sec:models} the repertoires, models, features, and file formats used in the literature are summarized. In Section~\ref{sec:authorship}, the existing problems of authorship attribution approached with computational style analysis are analyzed. Finally, in Section~\ref{sec:guidelines}, the literature is numerically analyzed and guidelines are drawn for future research in the composer identification task.

\section{Defining authorship attribution in MIR}
\label{sec:authorship-definition}

According to \cite{juola2007authorship}, \textit{authorship attribution} can be defined as "any attempt to infer the characteristics of the creator of a piece of [linguistic] data". In this context, "characteristics of the creator" refer to the stylistic and, in the original linguistic domain, textual patterns manifest in their work, which can be used for tasks such as identifying the author (attribution) or inferring other authorial traits (profiling). The problem is usually presented in one of three possible forms:
\begin{enumerate}
	\item In the \textit{closed-set} problem, the author is known to be one of a finite set of possible authors;
	\item In the \textit{open-set} problem, the author may not be one of the authors in the training set;
	\item In the \textit{profiling} problem, the author is unknown, and the task is to infer the author's characteristics.
\end{enumerate}

In the case of music composer identification, the literature does not report forms of type 2 and 3, so that the present work is mainly focused on problems of type 1.

In practice, in an authorship attribution task, a sample of data $D_\textit{unknown}$ with uncertain authorship is available. To infer the true attribution based on statistical analysis, researchers collect a dataset $D$ to train a model $M$ and apply it to $D_\text{unknown}$, obtaining predictions $P$. A key issue with this approach is determining how much trust can be placed in the predictions $P$.

Therefore, it is important to obtain a measure of merit for $M$, which serves as a measure of reliability for $P$. This remains an unsolved problem because the inherent complexity and high dimensionality of artistic style mean that the dataset used to train $M$, even if generally representative of the authors, frequently does not provide dense coverage for the specific region of the input space defined by a particular $D_\text{unknown}$.

A key challenge in applying composer identification techniques to musicological authorship attribution is distinguishing between features that genuinely reflect a composer's individual style and those that are merely indicative of broader categories such as historical period, genre, or nationality. While models trained to differentiate, for example, Baroque from Romantic composers might perform well for general classification or benchmarking purposes by leveraging these broad characteristics, their utility for fine-grained authorship attribution -- such as distinguishing between two contemporaneous composers of the same school -- is limited if they rely primarily on such extramusical factors rather than nuanced, individual stylistic traits. This survey critically examines the literature with this distinction in mind.

In the context of a typical machine learning experiment, the authorship attribution task has three key peculiarities:
\begin{enumerate}
	\item Researchers need to create a dataset $D$ that is as \textit{similar} as possible to the questionable data $D_\text{unknown}$ and should avoid introducing confounding variables or editorial choices that obscure the composer's intrinsic style. In the case of music, this means:
	      \begin{itemize}
		      \item defining the possible authors to be included in $D$;
		      \item collecting works and transcriptions that are musicologically valid and include minimal musicological interpretations or editorial choices;
		      \item encoding the music in a digital symbolic format that does not introduce confounding variables irrelevant to compositional style;
		      \item ensuring that the collected data encompasses all the composers' stylistic characteristics;
		      \item taking care of the quantitative aspects of the dataset, making it balanced with respect to styles and composers.
	      \end{itemize}
	\item The final inference on $D_\textit{unknown}$ is meaningless without a measure of merit computed on a sufficiently large evaluation dataset that differs from the one used to train the model $M$.
	\item In the \textit{open-set} problem, the evaluation of the model $M$ is particularly difficult, since the content of a class "anything else" is extremely difficult to define.
\end{enumerate}
Despite these peculiarities, the authorship attribution task can be considered a special case of the composer identification task.

While composer identification and authorship attribution can be applied to various media, this survey focuses on methods applied to symbolic music scores when the goal is to attribute authorship based on intrinsic compositional style. Consequently, I exclude studies that primarily rely on MIDI performances, audio recordings, or direct images of scores as their input. This exclusion is because these media types can introduce significant confounding variables that are distinct from the composer's original stylistic intent and can obscure the features relevant for musicological attribution:
\begin{itemize}
	\item MIDI performances are shaped by a performer's interpretation (e.g., dynamics, articulation, timing), which introduces variables related to performance choices rather than purely compositional ones. These interpretations might themselves be influenced by the performer's idea of the composer or prevailing cultural practices, further complicating the signal.
	\item Audio recordings are subject to performer interpretation, recording technology, room acoustics, and audio engineering choices, all of which can obscure or alter the purely compositional features.
	\item Images of music scores can introduce additional variables such as the quality of printing, specific editorial choices (especially in non-urtext editions), and the graphical style of particular editions, which are not direct reflections of the composer's intrinsic style.
\end{itemize}
Such confounding variables are not merely random noise but can introduce systematic patterns (e.g., a performer's consistent interpretation shaped by their understanding of the composer, or artifacts from a specific recording process or edition). If a model learns from these patterns, it might achieve high classification accuracy, but it becomes ambiguous whether the model is identifying the composer's intrinsic style or these external factors. This ambiguity undermines the goal of genuine authorship attribution based on compositional style, as the classification task itself may no longer reliably target the composer's unique stylistic traits. For this reason, the survey focuses on symbolic representations that aim to minimize these external influences.

For genres like popular music, "symbolic representation" for attribution purposes can encompass published scores, lead sheets, or other forms of notation that capture core compositional elements (melody, harmony, rhythm), provided the analysis focuses on these authorial aspects rather than performance or audio production characteristics. The key criterion for inclusion in this survey is that the input data aims to represent the composer's notated or structurally intended musical information, minimizing confounding variables from performance interpretation or audio production. Studies like \cite{glickman2019data}, which base their work on symbolic transcriptions of audio recordings to analyze compositional style, therefore fit within this scope.

For broader context and to delineate the survey's scope from its primary focus on composer attribution, I note that a related area involves works adopting computer vision techniques for attributing the scribe of a manuscript. While these work do not take into account the stylistic properties of the music, they are obviously of interest in the field of authorship attribution. Consequently, an additional search of documents was conducted and found 12 documents falling into this category. Of these, 4 make use of the dataset CVC-MUSCIMA \citep{fornes2012cvcmuscima}, which contains images of handwritten music collected in modern time, thus not focusing on historical composer identification problems. Other 4 studies are authored by Fornés \citep{fornes2008writer, fornes2009use, fornes2010combination, fornes2010symboldependent} and, as the remaining 4 articles \citep{bruder2004knowledgebased, gocke2003building, niitsuma2013writer, niitsuma2016musicologistdriven}, focus on historical manuscripts and set the purpose of their efforts in the automatic organization of digital archives. I found no real-world attribution problem that was tackled with computer vision techniques. The reason probably lies in the fact that manuscripts of the real composers are rarely available and, when they are, other more reliable techniques can be used for assessing the provenance of the manuscript.

\section{Collecting literature}
\label{sec:literature}
\subsection{Search methodology}
\label{sec:literature-methodology}

\begin{table*}[htbp]
	\center
	\begin{tabular}{|cc|c|}
		\hline \multicolumn{2}{|c|}{\textbf{Time blocks}} & \textbf{Search Key}                                                                        \\
		\textbf{Years}                                    & \multicolumn{1}{c|}{\textbf{Duration}} & \textbf{Token 1}                                  \\
		\hline 0-1950                                     & \multicolumn{1}{c|}{-}                 & \textbf{A} AND \textbf{B}                         \\
		\hline 1951-1970                                  & 20                                     & \textbf{A} AND \textbf{B}                         \\
		\hline \multirow{2}{*}{1971-1990}                 & \multirow{2}{*}{20}                    & \textbf{A} AND \textbf{B}                         \\
		                                                  &                                        & \textbf{A} AND \textbf{C}                         \\
		\hline 1991-2000                                  & 10                                     & \textbf{A} AND \textbf{C}                         \\
		\hline 2001-2010                                  & 10                                     & \textbf{A} AND \textbf{C} AND \textbf{D} -optical \\
		\hline 2011-2015                                  & 5                                      & \textbf{A} AND \textbf{C} AND \textbf{D} -optical \\
		\hline 2016-2018                                  & 3                                      & \textbf{A} AND \textbf{C} AND \textbf{D} -optical \\
		\hline 2019-2020                                  & 2                                      & \textbf{A} AND \textbf{C} AND \textbf{D} -optical \\
		\hline 2021-2023                                  & 2                                      & \textbf{A} AND \textbf{C} AND \textbf{D} -optical \\
		\hline 2024-                                      & 1                                      & \textbf{A} AND \textbf{C} AND \textbf{D} -optical \\
		\hline
	\end{tabular}
	\\
	\begin{tabular}{|c|c|}
		\hline \multicolumn{2}{|c|}{\textbf{Legend}}                                       \\
		\hline \textbf{A} & music AND composer AND style                                   \\
		\hline \textbf{B} & computer OR information OR statistics OR algorithm             \\
		\hline \textbf{C} & classification OR identification OR recognition OR attribution \\
		\hline \textbf{D} & "symbolic level" OR "music score" OR midi OR musicxml OR kern  \\
		\hline
	\end{tabular}
	\caption{
		\label{tab:queries}
		Queries used for each time block}
\end{table*}

The topic under investigation spans multiple decades during which the lexicon of computer science and statistics has evolved. Consequently, different queries were employed for distinct time periods. To account for this variation, time blocks were selected, and academic search engines were used to locate scientific papers in the field of composer identification from symbolic scores. In particular, the growing availability of digital academic documents in recent years necessitated the use of shorter time blocks for more recent periods. The selected time blocks are detailed in Table~\ref{tab:queries}.

A set of search queries was formulated using terminology adapted to different historical periods. These queries are listed in Table~\ref{tab:queries}. The terms "music," "composer," and "style" were retained consistently across all queries due to their specificity to the task. However, during the era of Youngblood's contributions, terms such as "classification"' or "recognition" were not associated with statistical learning methods. Consequently, more generic terms relevant to automation processes, such as "information," "statistics," "computer," and "algorithm," were used as substitutes.

Initially, computational limitations precluded large-scale digital analysis of sound signals. By the 2000s, however, the field of Music Information Processing (MIP) began to explore the audio domain extensively. Presently, the majority of literature in MIP focuses on audio signal processing methods. To refine the search results to studies utilizing the symbolic level, specific terms such as "symbolic level," "music score," "MIDI," "kern," and "musicXML" were included. As discussed in Section~\ref{sec:authorship-definition}, this focus is essential for investigating authorship attribution in music scores, which cannot be reliably applied to performance data or audio recordings due to potential confounding variables introduced by a performer’s stylistic interpretation, which are distinct from the composer's intrinsic style. Accordingly, studies employing audio recordings or MIDI performance data were excluded. A further refinement involved excluding works on Optical Music Recognition (OMR).

The search engine used was \textit{Google Scholar}, which retrieves results from major academic publishers and databases, including those indexed in \textit{Scopus} and \textit{Web of Science}, while extending the corpus to potentially include other documents. This engine’s capability to search within the full text of available works, rather than being limited to metadata such as titles and abstracts, was critical for identifying publications that, while not primarily focused on composer identification, still contained relevant information and experiments. Despite this advantage, the results often included noisy entries, and there was no assurance that all retrieved documents had undergone peer review. Only papers published in academic journals or presented at scientific conferences with a peer-review process involving at least two field experts were included. This criterion led to the exclusion of Bachelor’s, Master’s, and Ph.D. theses, as well as recent pre-print papers without peer review.

To further narrow the analysis, works that did not implement a full task of classifying musical compositions by composer were excluded. Specifically, papers providing only generic analysis without demonstrating the effectiveness of methodologies through classification tests with numerical results were not considered. However, studies utilizing both composers and genres in the classification labels were included, as such investigations are potentially relevant for characterizing composers' styles.

\begin{figure*}[htbp]
	\begin{center}
		\includegraphics[width=0.75\textwidth]{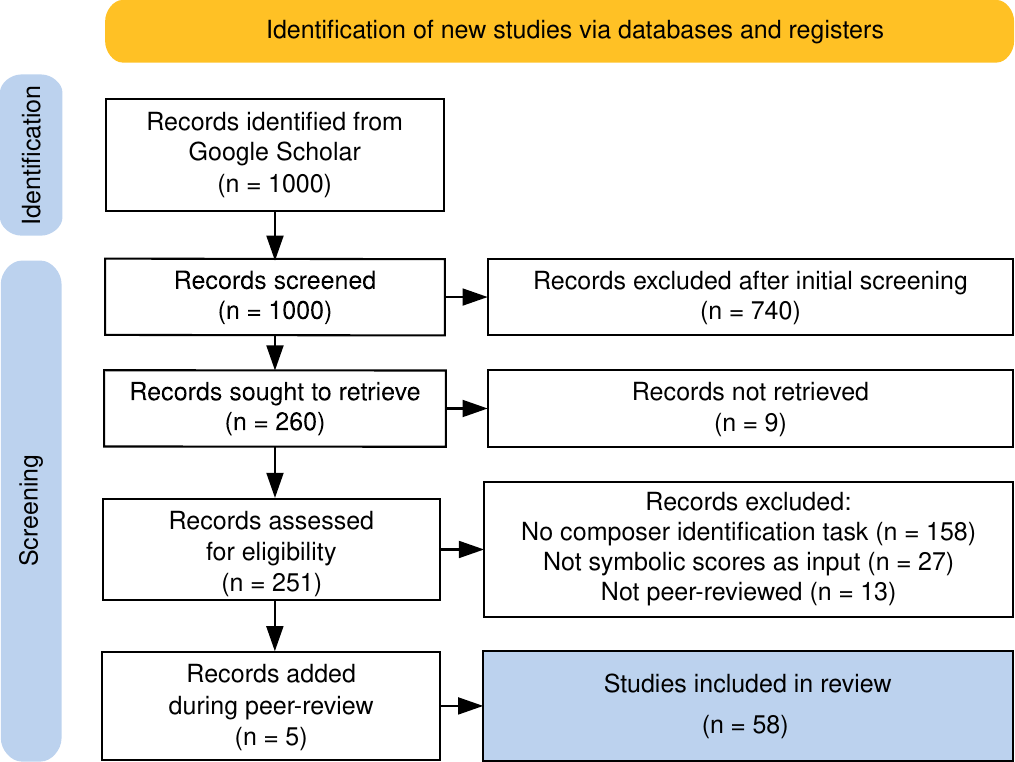}
	\end{center}
	\caption{PRISMA flowchart \citep{haddaway2022prisma2020} of the literature review conducted in this work.}
	\label{fig:prisma}
\end{figure*}

The first 10 pages of results from each of the 10 queries listed in Table~\ref{tab:queries} were analyzed, resulting in the consideration of 1,000 papers. From these, 53 papers meeting the specified criteria -- peer review, full classification tasks, and use of symbolic scores -- were selected. During the peer-review process, other 5 papers were added, primarily published during the review time-frame. The PRISMA flowchart summarizing the literature review process is presented in Fig.~\ref{fig:prisma}.

\subsection{Search results}
\label{sec:literature-results}

\begin{figure*}
	\begin{center}
		\includegraphics[width=\textwidth]{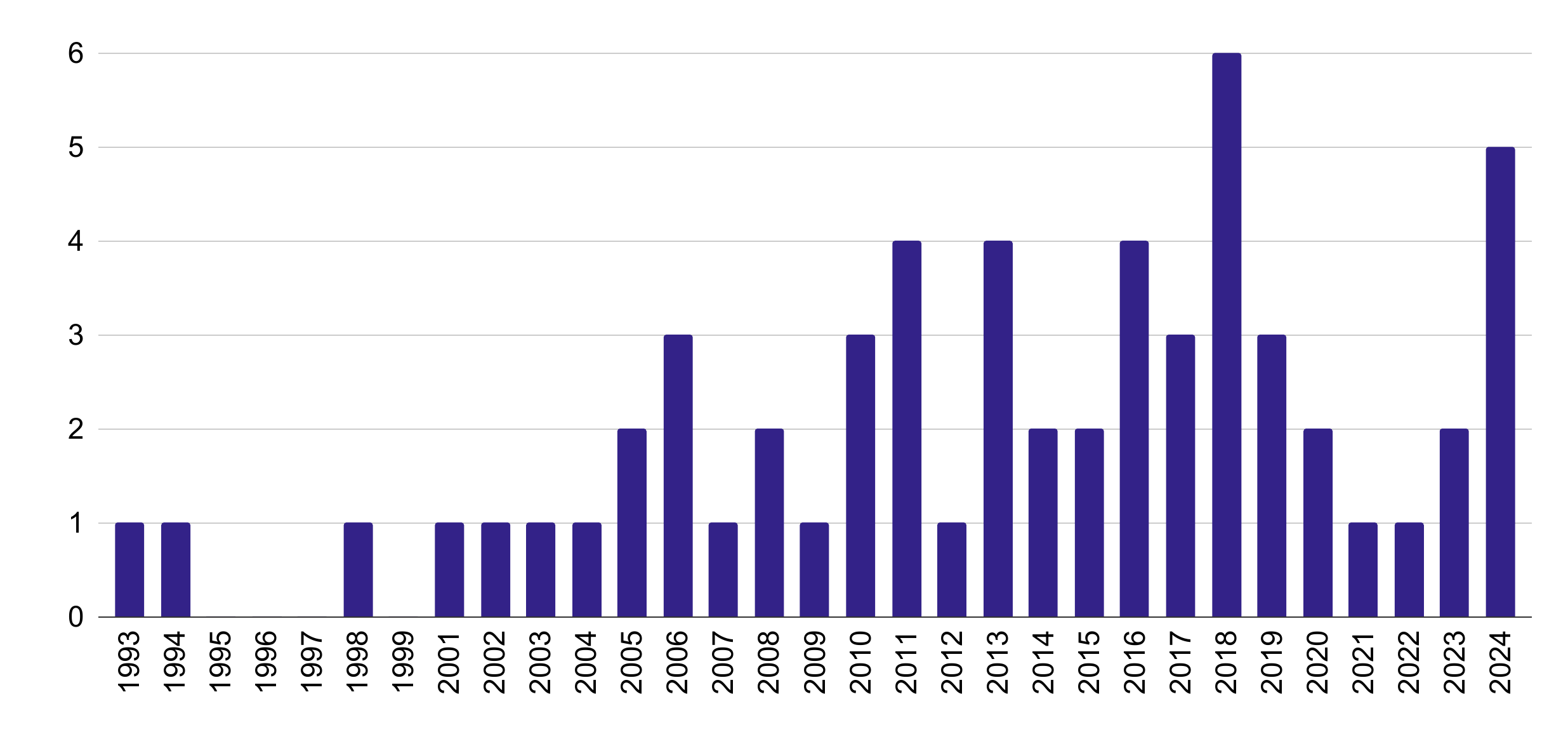}
	\end{center}
	\caption{Distribution of the number of publications across the years.}
	\label{fig:paper_year}
\end{figure*}

The statistical characterization of composers' styles has been studied since 1958 \citep{youngblood1958style}. However, the first work addressing the problem through the implementation of computational tools was published in 1993 at the Italian \textit{Colloquio di Informatica Musicale} \citep{johnson1993neural}. In the intervening period, particularly in the 1960s, several studies followed Youngblood's methodology by applying information theory to music analysis, but these did not compute proper classification performance \citep{fucks1962mathematical, fucks1962musical, gabura1965undergraduate, knopoff1983entropy, mendel1969preliminary, siromoney1964style} and generally did not utilize computational tools, with the exception of \cite{mendel1969preliminary}.

Two additional studies were published in the 1990s \citep{hornel1998learning, westhead1994automatic} prior to the work that is often cited as the earliest significant publication in the field \citep{pollastri2001classification}. Subsequently, 12 papers from the decade 2000–2009, 32 papers from the decade 2010–2019, and 11 papers since 2020 (excluding the present work) were identified. The distribution of these publications across the years is depicted in Figure~\ref{fig:paper_year}.

It should be noted that, since 2020, some studies have addressed composer identification using MIDI performance data \citep{chou2021midibertpiano, foscarin2022conceptbased, kong2020largescale, lee2020deep, yang2021composer}. These works were intentionally excluded to avoid confounding variables related to performer interpretation, as detailed in Section~\ref{sec:authorship-definition}.

In total, 58 works meeting the criteria outlined in Section~\ref{sec:literature-methodology} were collected. From each paper, all experiments involving composer identification tasks were reviewed, and the best-performing models for each dataset used in each paper were recorded. This process resulted in the identification of 87 models.

\section{Evaluation methodologies}
\label{sec:evaluation}

Given the critical importance of reliable evaluation protocols in authorship attribution \citep{rudman2012state}, it is essential to employ robust metrics that accurately reflect model performance.
A model with unsatisfactory performance metrics is unsuitable for authorship attribution because there its predictions are not reliable.
However, this principle is often overlooked in the literature, with some authors attributing works based on model parameters despite low classification effectiveness -- see Section~\ref{sec:authorship}.

Additionally, the reliability of the measure of merit strongly depends on the evaluation protocol, as \cite{sturm2014state} highlights: a well-designed protocol free from confounding variables is essential, and reproducibility requires selecting experimental components based on the scientific question rather than convenience. In any case, the evaluation measure is fundamental: if it is flawed, any conclusion drawn from it is questionable. To gain a comprehensive understanding of the literature, I categorized works based on their evaluation measures and analyzed other variables of the datasets for each category.

While this methodology may seem overly critical, it is essential to obtain a deeper understanding and define strong guidelines for future works regarding experimental design -- see Section~\ref{sec:guidelines}. It is also important to consider that the evaluation strategies in the surveyed literature may reflect the original aims of the studies. For instance, a paper focused on benchmarking a new algorithm might employ a dataset with clearly distinct composer styles and report $F_1$-measure, which, while potentially valid for its specific benchmarking goal, would be insufficient for assessing reliability in a nuanced musicological authorship attribution context. My critique primarily assesses evaluation choices from the perspective of their suitability for the latter.

While accuracy is a straightforward measure, it is often inadequate for imbalanced datasets, which are common in composer identification tasks. Despite its widespread use in the literature, accuracy can be misleading, as it does not account for the distribution of classes. The vast majority of the existing works in the analyzed literature adopt the standard accuracy, often as the only metric.

Some studies have utilized ROC curves, which plot the True Positive Rate (TPR) against the False Positive Rate (FPR) across different threshold settings. ROC curves provide qualitative insights into model performance, particularly in binary classification tasks. However, their applicability is limited when dealing with multiclass problems or when true labels for $D_\text{unknown}$ are unavailable. The Area Under the Curve (AUC) derived from ROC curves shares these limitations, making it less suitable for comprehensive evaluation. Four works \citep{glickman2019data,herremans2015classification,herremans2016composer,tan2019music} used AUC of the ROC.

The $F_1$-measure, originating from Information Retrieval, combines precision and recall into a single metric. While useful for highly imbalanced datasets, the $F_1$-measure can be problematic when the negative class is the minority, potentially leading to inflated scores that do not accurately reflect model performance. Additionally, its non-linear nature with respect to TPR and TNR can complicate interpretation \citep{christenReviewFmeasureIts2023}. In literature, 6 works \citep{hedges2014predicting,mckay2018jsymbolic,tan2019music,karystinaios2020musical,zhang2023symbolic,mirza2024decoding} adopted the $F_1$-measure in their analysis, while 1 work \citep{vannuss2017melody} used the AUC of the precision-recall curve.

Recent advances have highlighted the Matthews Correlation Coefficient (MCC) and Balanced Accuracy (BA) as more reliable alternatives \citep{Chicco2021TheMC}. Balanced Accuracy, in particular, has gained traction due to its simplicity, interpretability, and linearity with respect to TPR and TNR. This metric ensures balanced error rates across classes and is equivalent to standard accuracy when the dataset is perfectly balanced. For multiclass scenarios, Balanced Accuracy is extended by averaging the TPRs for all classes -- i.e. the average recall -- providing a fair evaluation across all categories and maintaining linear equivalence to the mean per class error and informedness. In the literature, 3 works explicitly adopted the BA \citep{simonetta2023optimizing,nakamura2015characteristics}, while other 2 reported BA under different names (``Overall Accuracy'' or ``Average Recall'') \citep{verma2019convolutional,mirza2024decoding}.

In this work, Balanced Accuracy is recommended as the preferred evaluation metric due to its robustness and interpretability, ensuring a more reliable assessment of model performance in the context of authorship attribution.
Moreover, it can be derived directly from the confusion matrix and is equivalent to standard accuracy when the classes are evenly distributed.
The BA of studies providing confusion matrices was recalculated for 10 works. Additionally, 19 works used the standard accuracy for datasets perfectly balanced. Overall, 33 works out of 58 were included in the analysis using BA as measure of merit.

Regarding the hold-out strategies, only 56 of the 88 recorded experiments use some form of cross-validation, of which 25 use a leave-one-out strategy. Of the remaining 32 experiments, 5 use random folds and 27 use a simple hold-out strategy.

The selected studies were categorized into three groups: "Unreliable," "Questionable Evaluation," and "Good Evaluation." The "Unreliable" category includes publications using inappropriate evaluation metrics and/or lacking cross-validation. The "Good Evaluation" category comprises studies where balanced accuracy could be computed and a $k$-fold cross-validation strategy was employed. The "Questionable Evaluation" category encompasses all remaining works, such as those using random fold cross-validation or relying on the $F_1$-measure.
It must be noted that, in those cases in which the dataset size and model complexity are really large, cross-validation procedures are hardly applicable.

\begin{figure*}[htbp]
	\begin{center}
		\includegraphics[width=\textwidth]{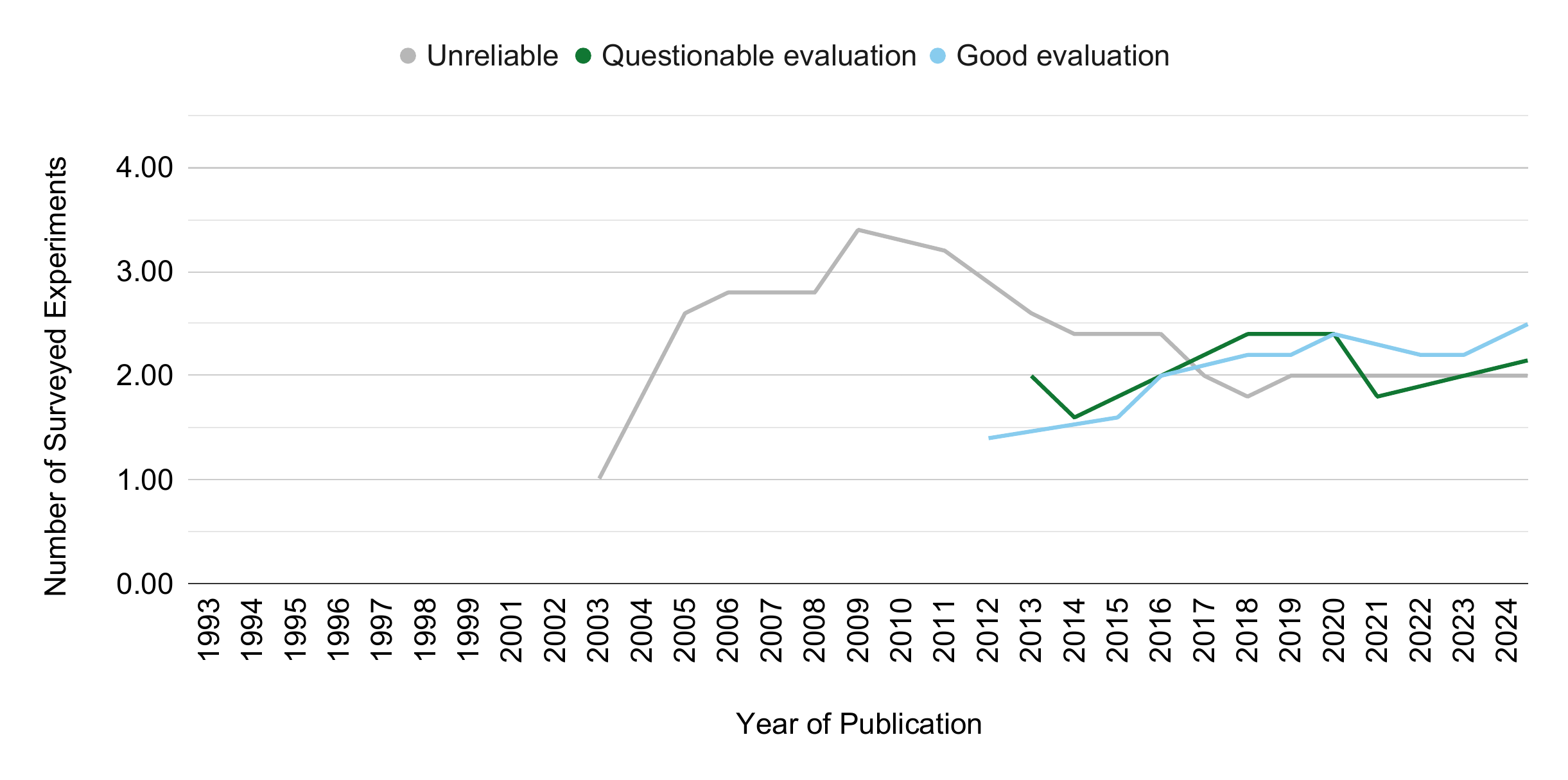}
	\end{center}
	\caption{The moving average on a window of 5 years of the number of surveyed experiments per year of publication across the evaluation categories identified in this survey.
	}
	\label{fig:experiments-year}
\end{figure*}

\begin{figure}[htbp]
	\begin{center}
		\includegraphics[width=0.5\textwidth]{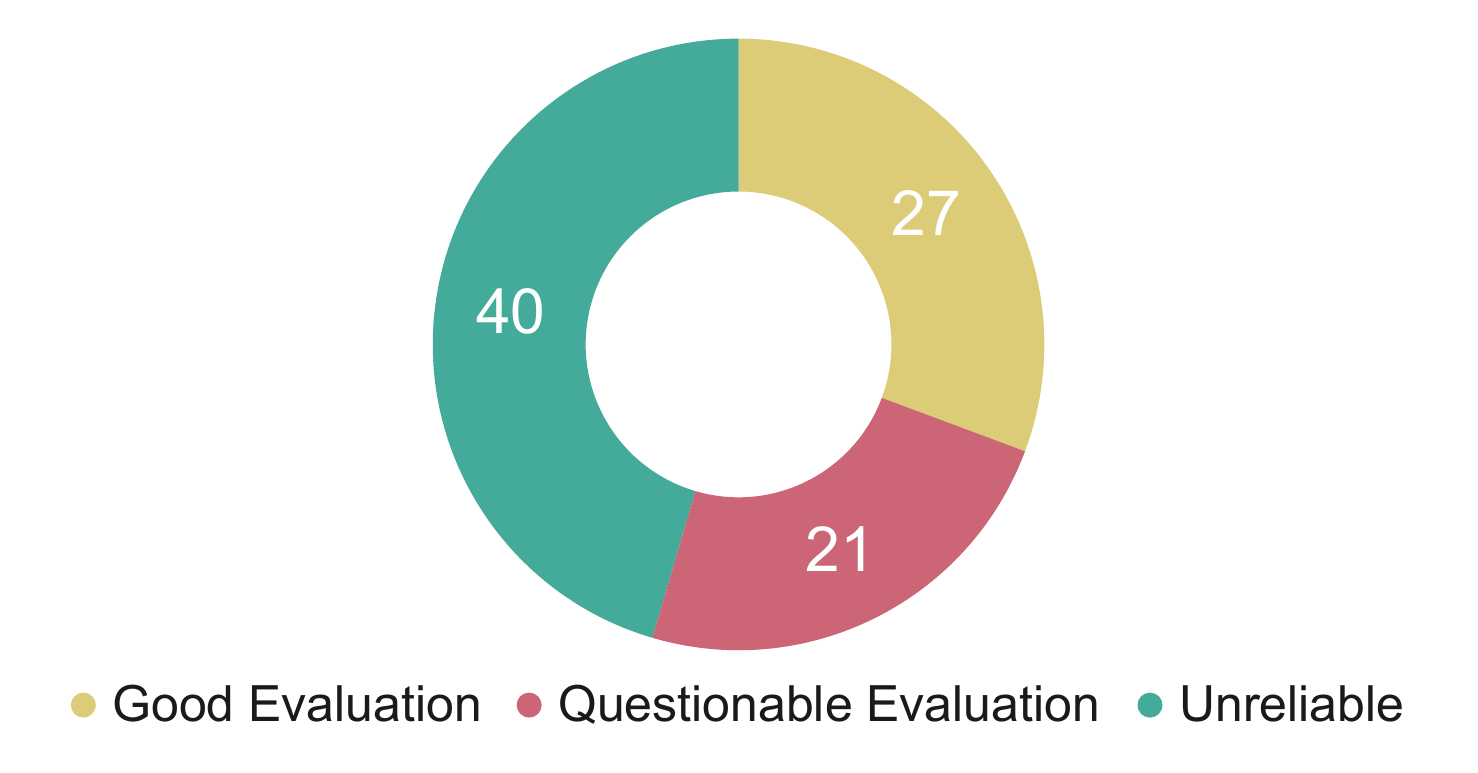}
	\end{center}
	\caption{Distribution of the recorded experiments across the evaluation classes identified.}
	\label{fig:evaluation}
\end{figure}

As illustrated in Figure~\ref{fig:evaluation}, 27 experiments were classified as having "Good Evaluation," 21 as "Questionable Evaluation," and 40 as "Unreliable" evaluation.
Additionally, Figure~\ref{fig:experiments-year} shows the number of experiments per year of publication across the three evaluation categories. The "Unreliable" category is prevalent in the earlier years, while the "Good Evaluation" category has gained prominence in recent years, particularly since 2010.

\section{Repertoire and datasets}
\label{sec:repertoire}

Some existing studies approach composer identification primarily as a classification benchmark rather than a musicological authorship attribution task. Particularly in earlier decades, computational models were frequently tested on heterogeneous repertoires, often spanning from the Baroque era to the 20th century, and in some instances, including medieval Gregorian chants \citep{cruz-alcazar2003musical}. Such material is typically selected because when the composers compared have markedly different cultural and aesthetic backgrounds, their musical styles can often be distinguished more easily through quantitative methods, simplifying the statistical challenge.
While this approach may be suitable for preliminary evaluations or for benchmarking tasks where the goal is to test a model's ability to capture gross stylistic differences, it is insufficient to demonstrate a model's capability to identify composers' styles accurately for the purpose of musicological attribution.
Effective evaluation for musicological authorship attribution must ensure the model is discriminating based on nuanced, individual musical style rather than easily separable extra musical factors that may correlate with style, such as genre, historical period, or geographical region \citep{rudman2012state}. Relying on these broader factors limits the model's applicability for distinguishing between, for example, contemporaneous composers within the same tradition or school.

Beyond the authorship attribution task, related MIR tasks include the automatic tagging of music in large databases \citep{vannuss2017melody} and comparative studies of composers' styles -- often focusing on Mozart and Haydn -- without addressing specific attribution problems \citep{backer2005musical,hillewaere2010string,hontanilla2011composer,wolkowicz2013evaluation,velarde2016composer,hajj2018automated,velarde2018convolutionbased,verma2019convolutional,kempfert2020where}. Even in these cases, the material selected for evaluation should not be distinguishable based on extra musical factors to ensure the validity of the results.

\begin{figure*}[htbp]
	\begin{center}
		\includegraphics[width=\textwidth]{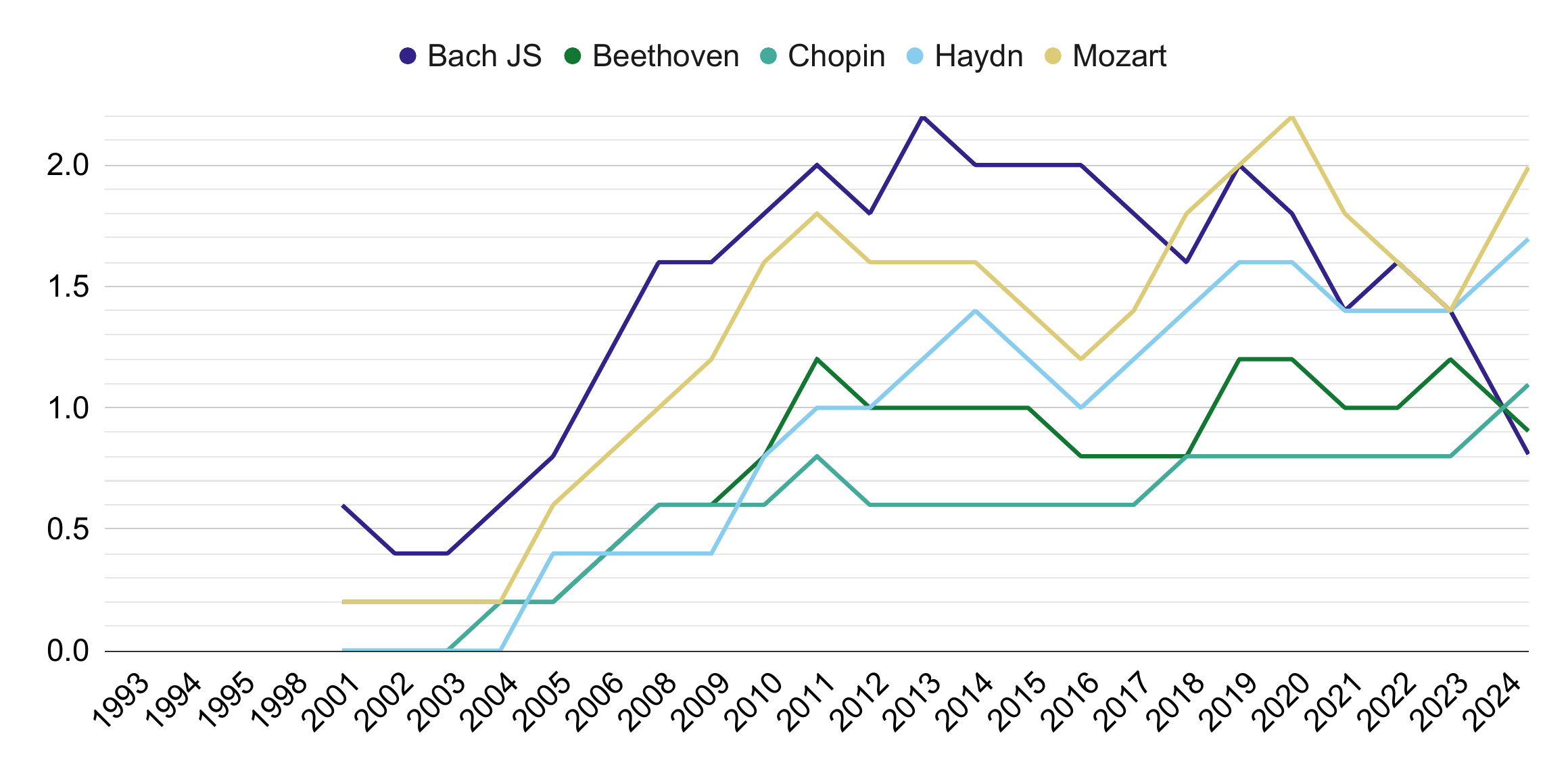}
	\end{center}
	\caption{
		The moving average on a window of 5 years of the number of papers per composer, for the 5 most common composers.
	}
	\label{fig:composers-time}
\end{figure*}

\begin{figure*}[htbp]
	\begin{center}
		\includegraphics[width=\textwidth]{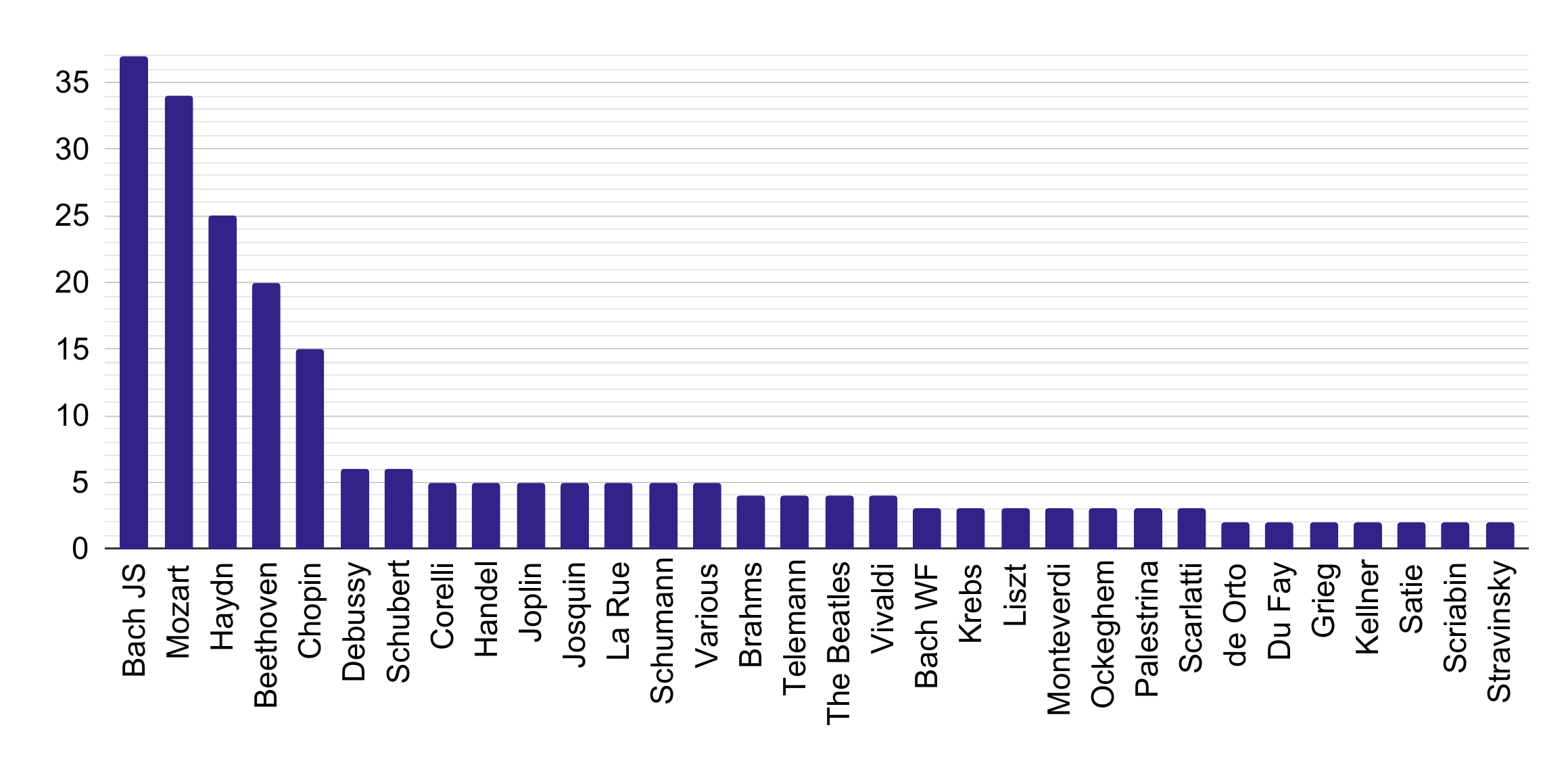}
	\end{center}
	\caption{Visualization of the number of papers for composers -- excluding composers for which only one paper was published.
	}
	\label{fig:composers-bars}
\end{figure*}

As shown in Fig.~\ref{fig:composers-bars}, the most frequently studied composer in composer identification tasks is Bach, followed by Mozart, Haydn, Beethoven, and Chopin. In recent years, there has been increased focus on Mozart, Haydn, and Beethoven, as illustrated in Fig.~\ref{fig:composers-time}. This distribution is likely influenced by the popularity of the CCARH corpora \citep{sapp2005online}, which are often utilized for music analysis in the symbolic domain.
Custom datasets have been rarely employed, and when used, they were typically unsupervised by musicologists, mined from user-contributed archives, and biased towards German composers and Western art music. Exceptions exist, primarily involving Chinese and Japanese folk music \citep{shan2002music,ruppin2006midi} and popular music, particularly The Beatles \citep{pollastri2001classification,glickman2019data}.

\begin{figure*}[tb]
	\begin{center}
		\includegraphics[width=0.75\textwidth]{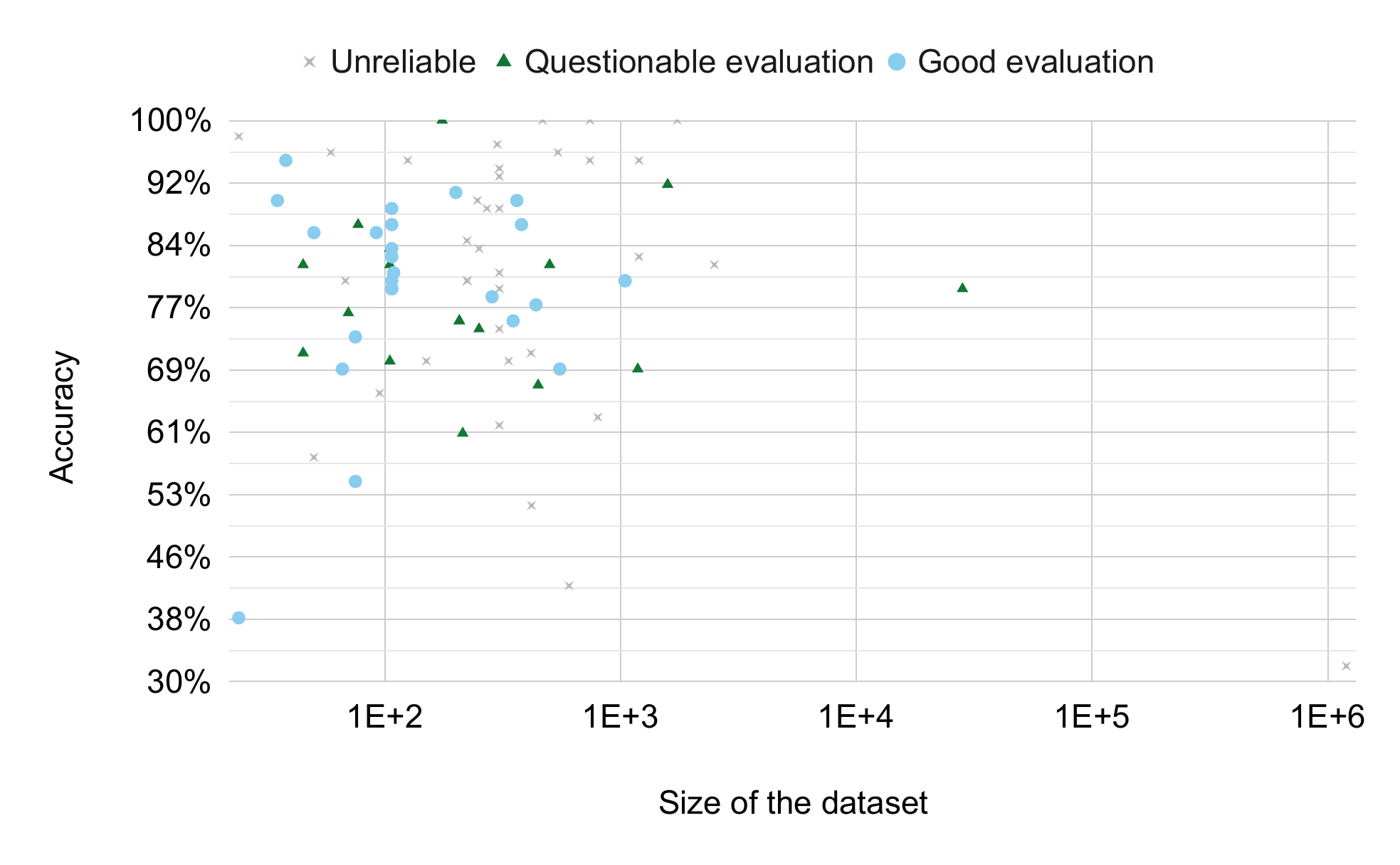}
	\end{center}
	\caption{
		The best accuracy reported by each paper across the size of the dataset. The classification "Unreliable"/"Questionable evaluation"/"Good evaluation" is made by the author based on the discussion in Section~\ref{sec:evaluation}.
	}
	\label{fig:accuracy-dataset}
\end{figure*}

The median dataset size in the reviewed experiments is 251 instances, with a minimum of 24 and a maximum of 1.2 million. 85\% of the experiments used a dataset with less than 1000 instances -- Figure~\ref{fig:accuracy-dataset} shows the distribution of the dataset size. However, comparing dataset sizes is challenging because individual instances may differ in duration, number of notes, measures, or movements. While this variation is less significant during the training phase, it is critical for the test phase, where true and false positives/negatives are counted in terms of instances.

The average number of classes in the datasets is 4, with a minimum of 2 and a maximum of 19, with some authors that did not report this aspect \citep{dor2011evaluation,vannuss2017melody}. For practical applications, the capacity to handle a large number of classes is generally advantageous but not essential if the composers are carefully selected by domain experts.

\begin{table*}[htbp]
	\begin{adjustbox}{rotate=90,max height=\textheight,center}
		\begin{tabular}{|llll|llll|}
			\hline
			Dataset            & Classes                                                   & Dataset Size         & Imbalance             & Paper                       & Accuracy & Cross-validation & Evaluation Class \\ \hline
			\multirow{3}{*}{1} & \multirow{3}{*}{Bach, Handel + Telemann + Haydn + Mozart} & \multirow{3}{*}{306} & \multirow{3}{*}{2.33} & Backer \& Kranenburg (2005) & 93\%     & LOO              & Unreliable       \\
			                   &                                                           &                      &                       & Backer \& Kranenburg (2005) & 94\%     & LOO              & Unreliable       \\
			                   &                                                           &                      &                       & Hontanilla et al. (2011)    & 89\%     & Not specified    & Unreliable       \\ \hline
			\multirow{4}{*}{2} & \multirow{4}{*}{Bach, Handel, Telemann, Haydn, Mozart}    & \multirow{4}{*}{306} & \multirow{4}{*}{1.74} & Backer \& Kranenburg (2005) & 74\%     & LOO              & Unreliable       \\
			                   &                                                           &                      &                       & Backer \& Kranenburg (2005) & 81\%     & LOO              & Unreliable       \\
			                   &                                                           &                      &                       & Hontanilla et al. (2011)    & 79\%     & Not specified    & Unreliable       \\
			                   &                                                           &                      &                       & Velarde et al. (2018)       & 62\%     & 5-folds          & Unreliable       \\ \hline
			\multirow{7}{*}{3} & \multirow{7}{*}{Haydn, Mozart}                            & \multirow{7}{*}{107} & \multirow{7}{*}{1.02} & Backer \& Kranenburg (2005) & 79\%     & LOO              & Good             \\
			                   &                                                           &                      &                       & Velarde et al. (2016)       & 79\%     & LOO              & Good             \\
			                   &                                                           &                      &                       & Velarde et al. (2018)       & 80\%     & LOO              & Good             \\
			                   &                                                           &                      &                       & Kempfert \& Wong. (2020)    & 84\%     & LOO              & Good             \\
			                   &                                                           &                      &                       & Takamoto et al. (2024)      & 83\%     & LOO              & Good             \\
			                   &                                                           &                      &                       & Alvarez et al. (2024)       & 87\%     & LOO              & Good             \\
			                   &                                                           &                      &                       & Gelbukh et al. (2024)       & 89\%     & LOO              & Good             \\ \hline
			\multirow{3}{*}{4} & \multirow{3}{*}{Haydn, Mozart}                            & \multirow{3}{*}{207} & \multirow{3}{*}{1.18} & Hillawaere et al. (2010)    & 75\%     & LOO              & Questionable     \\
			                   &                                                           &                      &                       & Wołkowicz \& Kešelj (2013)  & 75\%     & LOO              & Questionable     \\
			                   &                                                           &                      &                       & Velarde et al. (2018)       & 75\%     & LOO              & Questionable     \\ \hline
		\end{tabular}
	\end{adjustbox}
	\caption{Datasets used by at least 3 studies and related accuracies. The "Imbalance" has been computed as the ratio between the largest and the smallest class. }
	\label{tab:datasets}
\end{table*}

Only a few datasets have been used in more than three studies. Their details and associated accuracies are summarized in Table~\ref{tab:datasets}.

\begin{figure*}[tb]
	\begin{center}
		\includegraphics[width=0.75\textwidth]{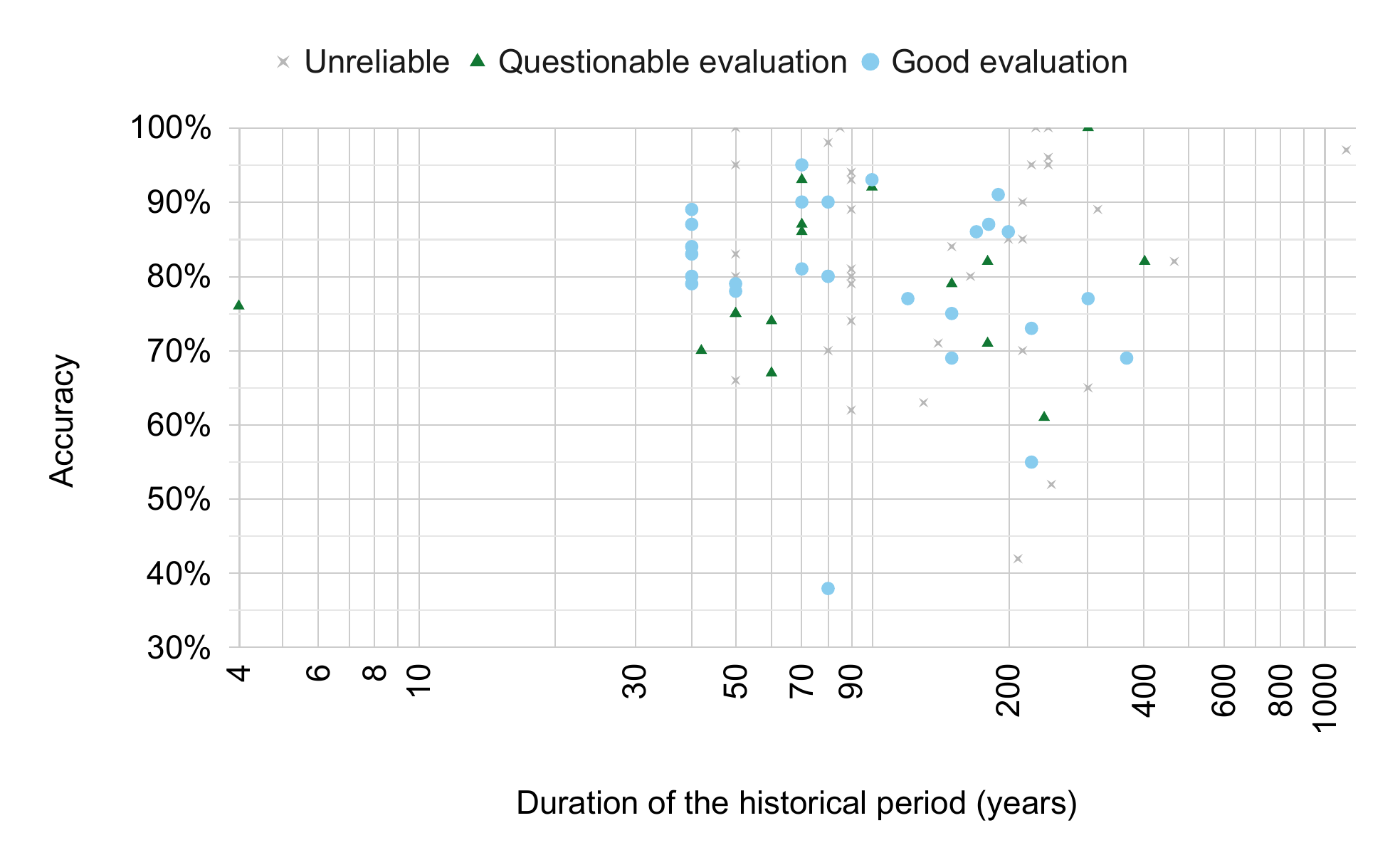}
	\end{center}
	\caption{
		The best accuracy reported by each paper across the length of the period considered by the respective paper. The classification "Unreliable"/"Questionable evaluation"/"Good evaluation" is made by the author based on the discussion in Section~\ref{sec:evaluation}.
	}
	\label{fig:accuracy-period}
\end{figure*}

I attempted to evaluate the model accuracy -- or balanced accuracy where it could be computed -- based on the type of dataset. The resulting scatter plots are shown in Figures~\ref{fig:accuracy-period} and~\ref{fig:accuracy-dataset}.

\section{Representational Frameworks for Symbolic Music}
\label{sec:representations}

The way symbolic music is represented significantly impacts the effectiveness of composer identification models. The literature showcases a variety of approaches, which can be broadly categorized by how features are derived and their fundamental representational form, as detailed below. These representations are typically derived from various digital music file formats (see Section~\ref{subsec:source_file_formats}).

\subsection{Feature representations}

Different feature representations emphasize distinct aspects of musical structure, influencing both the interpretability and effectiveness of computational models. This section outlines the primary categories of feature representations, their construction, relevance, and applications.

\subsubsection{Statistically-derived descriptors}
Statistically-derived descriptors summarize musical content using statistical measures or counts of predefined elements. These features include frequencies of pitches, intervals, chords, or rhythmic values, as well as higher-order statistics like means, variances, entropy, and Zipf-Mandelbrot parameters \citep{marques2013recognizing, machado2004adaptive}. Tools such as jSymbolic \citep{mckay2018jsymbolic}, Music21 \citep{cuthbert2011feature}, and Musif \citep{llorens2023musif} automate the extraction of these descriptors. A recent comparative study by \cite{simonetta2023optimizing} highlights that the choice and combination of statistically-derived descriptors are crucial for effective composer identification. Their findings indicate that the optimal feature set is often task-dependent -- meaning that different musical characteristics become more or less salient depending on the specific analytical goal (e.g., broad period classification versus distinguishing closely related composers). Furthermore, the study suggests potential benefits from combining features derived from different extraction tools or methodologies, as these can capture complementary facets of musical style, leading to a more comprehensive representation.
This type of descriptors can capture pitch-class distributions, harmonic dissonance, or adherence to compositional rules \citep{doerfler2013approach, johnson1993neural}. These features are typically represented as scalar values or numerical vectors, such as histograms or frequency distributions, and are widely used with machine learning models like SVMs, Naive Bayes, Decision Trees, and k-NN. Their strengths lie in their ability to adeptly summarize overall musical characteristics into compressed, often interpretable values, and their computational efficiency; however, they typically abstract away temporal information and may require extensive hand-crafting. Despite these limitations, they remain effective for tasks like composer classification, music similarity, and corpus analysis \citep{backer2005musical, sadeghian2017classification}.

\subsubsection{Sequential and pattern-based descriptors}
Sequential and pattern-based descriptors explicitly model the order and recurrence of musical events. N-grams, motif mining, and grammar-based approaches are common in this category. N-grams extract contiguous subsequences of $n$ items (e.g., notes, intervals, chords) and are often paired with frequency counts, TF-IDF weighting, or probabilistic models like Markov chains \citep{alvarez2024composer, hontanilla2011composer, wolkowicz2008ngrambased}. Motif mining identifies repeated subsequences, while grammar-based methods infer hierarchical structures using formal grammars \citep{cruz-alcazar2003musical, mondol2021grammarbased}. Compression-based techniques, which rely on repetitive patterns, also fall within this category \citep{junior2012composer, ruppin2006midi}. Recent tokenization schemes, such as REMI \citep{huang2020pop} and Compound Word  \citep{hsiao2021compound}, encode multiple musical aspects into single tokens, facilitating sequential modeling. These representations, whether as token sequences, frequency vectors, or grammar-derived structures, are particularly effective for capturing temporal dependencies, melodic contours, and harmonic progressions. Yet, significant challenges persist: sparsity, particularly for high-order $n$-grams, inherently limits generalization and scalability; tokenization choices critically affect all symbolic sequential methods, trading off information capture and vocabulary size; the local focus of $n$-grams struggles with music's long-range dependencies, which computationally expensive and potentially ambiguous grammar induction addresses only partially; the rigidity to variations in exact $n$-gram and motif matching hampers robustness against minor musical alterations; and high computational demands, especially for uxhaustive motif mining and complex grammar induction, restrict large-scale applications. These factors necessitate careful design and often motivate hybrid approaches. Nevertheless, sequential and pattern-based descriptors remain widely applied in classification, melody retrieval, music generation, and structural analysis.

\subsubsection{Transformational and geometric representations}
Transformational and geometric representations map symbolic music into alternative spaces where patterns may be more apparent or amenable to specific processing techniques. Examples include piano-roll matrices, which represent pitch versus time as binary or valued grids \citep{velarde2016composer, zhang2023symbolic}; spectrogram-like representations derived from symbolic data via synthesis and time-frequency transforms~\citep{velarde2018convolutionbased, }; geometric spaces like the Tonnetz, which map harmonic relationships onto a lattice \citep{karystinaios2020musical}; and wavelet or time-frequency transforms applied to musical parameters \citep{mirza2024decoding}. These formats are particularly suited for convolutional neural networks, which can exploit spatial locality to identify structural patterns. While these representations can reveal holistic or multi-scale patterns, they arrepresentation often high-dimensional. Parameterization, such as defining time/pitch resolution for piano-rolls or selecting parameters for wavelet transforms (e.g., \citep{velarde2016composer, mirza2024decoding}), can be critical and extensive, shaping the core representation more significantly than the selection process for many statistical descriptors or the use of fixed pre-trained learned representations. Regarding interpretability, while base forms like piano-rolls or the Tonnetz lattice (e.g., \citep{karystinaios2020musical}) are conceptually intuitive, they are used in literature by \textit{deriving} features or transformed representations for modeling, obscuring direct musical elements. They consequently offer less direct linkage to specific musical meaning compared to features from statistical or simple sequential approaches. They are primarily used for classification, segmentation, and visualization.

\subsubsection{Graph-based representations}
Graph-based representations conceptualize music as networks of interconnected entities, such as notes or chords, with edges encoding relationships like succession, simultaneity, or voice leading \citep{karystinaios2024perceptioninspired, zhang2023symbolic}. Features can be derived from node or edge attributes, global graph statistics (e.g., centrality, clustering), or learned embeddings from graph neural networks. Some approaches extend this paradigm to model higher-level structures, such as entire MIDI files and their components, as nodes in a graph \citep{lisena2022midi2vec}. Graph-based representations excel at capturing complex, non-local dependencies and structural properties, making them powerful for tasks requiring a nuanced understanding of musical relationships. However, designing graphs that meaningfully capture music's multi-dimensional, concurrent, and often non-linear relational structure -- defining appropriate nodes, varied edge types, and attributes to form an explicit relational representation -- is an intricate process, entailing a more multifaceted design of relational structures than defining linear token sequences or statistical aggregates. While GNNs can be parameter-efficient \citep{zhang2023symbolic}, their computational aspects, such as processing time for message passing over large or intricate musical graphs and the potential for extended convergence times \citep{zhang2023symbolic, karystinaios2024perceptioninspired}, warrant careful consideration, especially as graph size and complexity scale.

\subsubsection{Learned and latent representations}
Learned and latent representations leverage deep learning models to automatically infer relevant features from raw or other primary symbolic representations. These representations correspond to activations of hidden or embedding layers within neural networks—such as CNNs, RNNs, Transformers, or autoencoders—trained on musical tasks, or arise from graph embedding techniques \citep{lisena2022midi2vec, karystinaios2024perceptioninspired}. While other methods produce intermediate results, these learned features are distinct as they are optimized by the model and designed to be reusable across diverse tasks. Crucially, such pre-computed representations can potentially be adopted by other researchers for novel applications, potentially without the need for complete retraining from scratch. For example, MIDI2vec embeddings are used for predicting various metadata \citep{lisena2022midi2vec}, and perception-inspired graph convolutions generate hidden representations applicable to multiple music understanding problems \citep{karystinaios2024perceptioninspired}. Learned representations are highly expressive and can capture subtle characteristics that elude manual feature engineering. However, they are generally opaque, require large datasets and computational resources, and their effectiveness depends on the quality and diversity of training data.

\subsection{Source File Formats}
\label{subsec:source_file_formats}
The aforementioned representations are derived from various digital symbolic music file formats. The choice of format can influence the type and richness of information available for feature extraction or representation learning.
\begin{itemize}
	\item Standard MIDI File (SMF): Widely adopted due to its simplicity and broad software support. It primarily encodes performance-oriented data like pitch, duration (as onset and offset or duration), and velocity. While excellent for capturing basic note events, it often lacks detailed notational information like precise beaming, enharmonics, or high-level structural annotations, which can be crucial for some stylistic analyses.a Example of studies that SMF are \cite{chan2006recognition,costantini2006automatic,cumming2018methodologies, hajj2018automated,simonetta2023optimizing}.
	\item **kern: It is effective for representing polyphony in Western classical music and can encode a wide range of musicological details. It is relatively human-readable but can be complex to create manually. Example of works using **kern format are \cite{alvarez2024composer,brinkman2016musical,hillewaere2010string,simonetta2023optimizing}.
	\item MusicXML: A de facto standard for exchanging modern music notation between different software. It is more comprehensive than MIDI, capable of representing detailed notational elements like slurs, dynamics, and articulations. However, its verbosity can be a challenge, and it primarily focuses on common Western music notation. Example of articles that used MusicXML are \cite{doerfler2013approach, simonetta2023optimizing}.
\end{itemize}
Other formats like ABC notation and Guido Music Notation also exist, often prioritizing human readability or specific notational capabilities, but are less frequently encountered as primary sources in the surveyed articles. A major effort is being undertaken by the Music Encoding Initiative (MEI), an XML framework designed for encoding a wide variety of music notations with rich metadata, including editorial and critical annotations, with a special focus on scholarly applications.

Interestingly, studies by \cite{simonetta2023optimizing} and \cite{zhang2023symbolic} have found that for certain composer identification tasks, using highly informative file formats like MusicXML or MEI did not always yield significantly better results than simpler formats like MIDI when using current feature extraction tools or neural representation methods. This may indicate that existing tools and models do not yet fully exploit the rich semantic information these formats provide. Traditional feature extraction often relies on aspects readily available or derivable from MIDI-like data, and many neural models are designed to learn features from more fundamental representations. This suggests an opportunity for future research to develop methods that can more effectively leverage the detailed musicological information captured in comprehensive encoding formats for improved stylistic analysis.

\section{Model Architectures}
\label{sec:models}

\begin{figure*}[htbp]
	\begin{center}
		\includegraphics[width=0.75\textwidth]{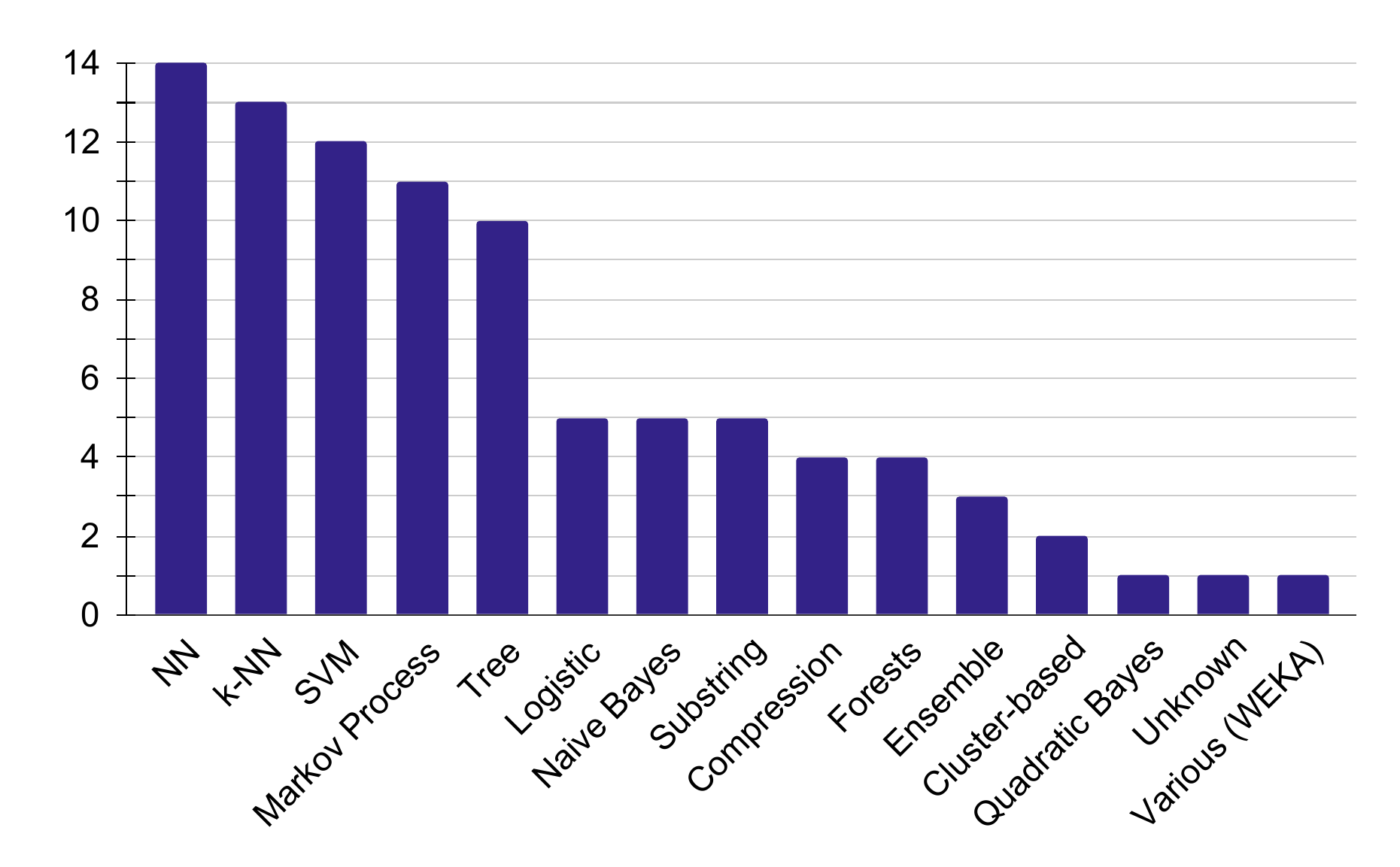} 
	\end{center}
	\caption{Visualization of the number of papers for each approach.}
	\label{fig:methods-bars}
\end{figure*}

\begin{figure*}[htbp]
	\begin{center}
		\includegraphics[width=\textwidth]{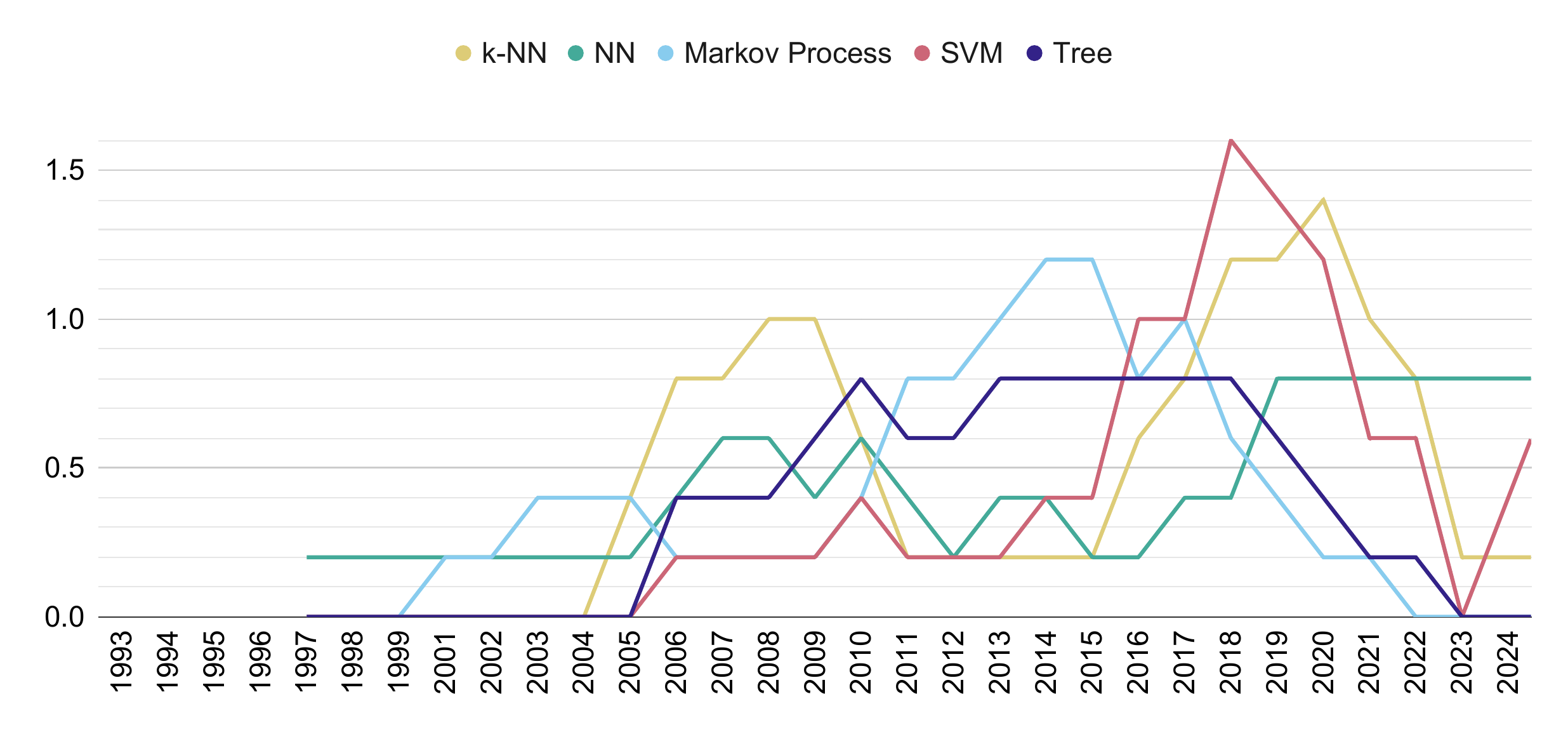} 
	\end{center}
	\caption{
		The moving average on a window of 5 years of the number of papers per approach, for the 5 most common approaches.
	}
	\label{fig:methods-time}
\end{figure*}

The task of composer identification has been approached with various models, each leveraging different strengths in statistical learning. This section categorizes these models based on their underlying architecture and discusses their suitability for different types of feature representations, as detailed in Section~\ref{sec:representations}. Figure~\ref{fig:methods-bars} shows the distribution of papers employing each approach.

To create these statistics, base classes of approaches were identified, and each recorded experiment was associated with a specific class. For instance, decision trees fall into the class of "Tree", random forests into the class of "Forests", and the same applies to convolutional neural networks, transformers, and feed-forward neural networks, all classified as "NN". Similarly, "Markov Processes" are used for any methodology that leverages the theory of Markov processes, including HMM and simple Markov chains.

\subsection{$k$-Nearest Neighbors ($k$-NN)}
$k$-NN is a frequently used, simple approach in music classification. These models operate directly on feature vectors, where each dimension represents a specific musical characteristic. The representation space can be formed
statistically-derived descriptors (e.g., \citep{backer2005musical, brinkman2016musical}),  transformational and geometric representations like piano-roll \citep{velarde2018convolutionbased} and Tonnentz \citep{karystinaios2020musical}
and ad-hoc bit-level representations suitable for compression-based distances \citep{takamoto2016improving}, which relate to sequential and pattern-based descriptors \citep{hajj2018automated, ruppin2006midi}.
The drawbacks of $k$-NN are its poor interpretability and its increasing computational cost with larger training sets.

\subsection{Neural Networks (NNs)}
Neural networks encompass a broad range of architectures, each suited to different input representations:
\begin{itemize}
	\item Feed-forward NNs \citep{johnson1993neural,hornel1998learning,machado2004adaptive,costantini2006automatic,kaliakatsos-papakostas2010musical,marques2013recognizing,tan2019music,lisena2022midi2vec,sadeghian2017classification}: These typically take fixed-size feature vectors as input, making them compatible with statistically-derived descriptors.
	\item Convolutional Neural Networks (CNNs) \citep{verma2019convolutional,zhang2023symbolic,velarde2018convolutionbased}: CNNs are often applied to matrix-based representations like piano rolls, aligning with transformational and geometric representations. They can learn hierarchical patterns across time and pitch.
	\item Recurrent Neural Networks (RNNs) and Transformer models \citep{chou2021midibertpiano, zhang2023symbolic}: These are well-suited for sequence-based representations, such as streams of musical events (e.g., notes, rests, tokens representing pitch and duration). Transformers are one of the most used architecture typology in modern NN, but they suffer from memory limitations when processing long sequences and generally require large datasets to train effectively.
	\item Graph Neural Networks (GNNs) \citep{zhang2023symbolic,karystinaios2024perceptioninspired}: GNNs operate on graph structures representing relationships between musical elements, directly leveraging graph-based representations.
\end{itemize}
Other variants like fuzzy neural networks \citep{sadeghian2017classification} and cortical algorithms \citep{hajj2018automated} also exist. Compared to trends in machine learning, where NNs are increasingly prevalent, their application in composer identification remains relatively less explored, potentially due to the limited size of curated datasets.

\subsection{Markov Processes}
Markov processes are based on the assumption that music is a Markov process, where the probability of a musical symbol depends solely on the preceding one. This approach aligns with sequential and pattern-based descriptors. The musical symbol can vary but often includes discrete elements like pitches, intervals, durations, or combinations thereof. This category includes
Markov Chains \citep{kaliakatsos-papakostas2011weighted,nakamura2015characteristics,neto2017methods},
$n$-grams \citep{cruz-alcazar2003musical,wolkowicz2008ngrambased,hillewaere2010string,hontanilla2011composer,hontanilla2013modeling,wolkowicz2013evaluation,hedges2014predicting,alvarez2024composer,ruppin2006midi,shan2002music}, and
Hidden Markov Models (HMMs) \citep{pollastri2001classification}.
Markov models often struggle with representing music polyphony. Methods dealing with polyphonic music range from considering only the highest part, simplifying it and automatically extracting the melody \citep{simonetta2019convolutional}, to concatenating the various parts sequentially, and modeling each part with a different model. As shown in Fig.~\ref{fig:methods-time}, the use of Markov Processes has declined in recent times, likely being replaced by NNs with LSTM or Attention methods \citep{chou2021midibertpiano, zhang2023symbolic}.

\subsection{Support Vector Machines (SVMs)}
SVMs apply a non-linear transformation to the data, seeking a space where the data are linearly separable. SVMs typically operate on feature vectors derived from various musical aspects, such as statistical summaries of melody, harmony, and rhythm, often extracted using tools discussed in Section~\ref{sec:representations}. Thus, they are well suited for statistically-derived descriptors. SVMs have been used both in comparison to other methods \citep{costantini2006automatic,hillewaere2010string,herremans2015classification,hajj2018automated,velarde2018convolutionbased,herlands2014machine,brinkman2016musical,chan2006recognition,naccache2009learningbased}, and as a sole choice \citep{velarde2016composer,mckay2018jsymbolic,alvarez2024composer,gelbukh2024multiinstrument}.

\subsection{Decision Trees and Random Forests}
Ten works utilized Decision Trees -- including RIPPER -- primarily as an option to interpret the model decision \citep{chan2006recognition,costantini2006automatic,naccache2009learningbased,mearns2010characterisation,cuthbert2011feature,marques2013recognizing,herlands2014machine,herremans2015classification,herremans2016composer,costa2018characterization,dor2011evaluation}. These also operate on feature vectors, making them suitable for statistically-derived descriptors. Their use is controversial, as they rarely achieve high accuracies but are a good choice for interpretability. Random Forests, an ensemble of decision trees, provide a more effective framework that is more commonly used in machine learning, albeit with decreased model interpretability, and have been adopted by only three works in the context of composer identification \citep{herlands2014machine,tan2019music,karystinaios2020musical}.

\subsection{Other Methods}
Other methods found in the literature include Naive Bayes, mainly adopted as a baseline \citep{chan2006recognition,costantini2006automatic,naccache2009learningbased,mearns2010characterisation,herlands2014machine,dor2011evaluation}, which also uses feature vectors; compression-based distances used for clustering or $k$-NN \citep{ruppin2006midi,junior2012composer,takamoto2016improving,mondol2021grammarbased}, which operate on sequential representations; Quadratic Bayes \citep{backer2005musical}; and ad-hoc clustering \citep{doerfler2013approach}. Substring matching was adopted by four works, operating on sequences of musical symbols \citep{hedges2014predicting, shan2002music, westhead1994automatic,takamoto2024composer}. This method is based on the classification of a test sequence with the label whose training set has the highest number of sub-sequences shared with the test instance. Some authors utilized the WEKA framework to compare a variety of methods \citep{dor2011evaluation,simonetta2023optimizing}, while more recently, AutoML methods have been successfully used with surprising efficacy \citep{simonetta2023optimizing}.

The methods analyzed in literature were not used homogeneously across time, but followed trends of the general machine learning field. Figure~\ref{fig:methods-time} shows the trend of the five most common approaches.

\section{Authorship attribution problems}
\label{sec:authorship}

\begin{table*}[htbp]
	\centering
	\begin{adjustbox}{max width=\textwidth}
		\begin{tabular}{|m{4cm}|m{4.5cm}|m{4.5cm}|m{4.5cm}|}
			\hline
			\textbf{Disputed attribution} & \textbf{Bach?}                                                                  & \textbf{Josquin Desprez?}                                           & \textbf{Lennon-McCartney}                                                                          \\ \hline
			\textbf{Originating work}     & \cite{vankranenburg2005musical}                                                 & \cite{brinkman2016musical}                                          & \cite{glickman2019data}                                                                            \\ \hline
			\textbf{Number of papers}     & 3                                                                               & 3                                                                   & 1                                                                                                  \\ \hline
			\textbf{Best accuracy}        & >90\% \citep{vankranenburg2008measuring}                                        & 91\% \citep{mckay2018jsymbolic}                                     & 76\% \citep{glickman2019data}                                                                      \\ \hline
			\textbf{Evaluation class}     & Good                                                                            & Questionable                                                        & Questionable                                                                                       \\ \hline
			\textbf{Data integrity}       & Closed set, reliance on CCARH collection, no investigation on editorial choices & Closed set, reliance on JRP collection, weak transcription protocol & Closed set, manual transcription based on previous editions, no investigation on editorial choices \\ \hline
		\end{tabular}
	\end{adjustbox}
	\caption{Summary of music authorship attribution problems in the analyzed literature.}
	\label{tab:authorship_attribution}
\end{table*}

Only three cases of unknown or questionable authorship have been approached with composer identification models. The first case involves a set of keyboard fugues originally attributed to Johann Sebastian Bach, but recently re-attributed to Krebs, supported by various statistical analyses \citep{vankranenburg2005musical,vankranenburg2008measuring,hontanilla2011composer}. The second case pertains to more than 200 works of questionable attribution by Josquin Desprez \citep{brinkman2016musical,mckay2018jsymbolic,verma2019convolutional}. Finally, the third case concerns 8 songs (or portions of songs) with disputed attribution to John Lennon and Paul McCartney \citep{glickman2019data}.

In the Bach fugue problem, the initial work \citep{vankranenburg2005musical} identified a set of possible alternative authors based on historical observations and musicological considerations, thus restricting the inference problem to three authors: Johann Sebastian Bach, Wilhelm Friedemann Bach, and Johann Ludwig Krebs. The authors collected music scores from the CCARH collection \citep{sapp2005online}. The same authors later included Johann Peter Kellner as an option in a more extended analysis \citep{vankranenburg2008measuring}. Other researchers approached the problem in the second formulation, obtaining similar results \citep{hontanilla2011composer}. It should be noted that the full text of the latter work was not accessible, and the analysis was based on the conference slides made available by the authors and on a subsequent publication describing the same method in full detail but excluding the authorship attribution problem \citep{hontanilla2013modeling}. Upon examining the evaluation procedures of these three works, it was found that the first work did not adopt proper evaluation measures for the imbalanced dataset at hand, while the third work showed poor classification performance, thus undermining the reliability of the model. The second work \citep{vankranenburg2005musical}, however, allowed for proper evaluations based on the confusion matrix reported in the original publication, showing a balanced accuracy larger than 90\% in a leave-one-out evaluation procedure. Therefore, it is suggested that this latter work \citep{vankranenburg2008measuring} contains the most definitive evaluation up to now. Regarding the construction of the dataset, not much attention was paid by the authors on the elimination of potential confounding factors stemming from editorial choices in the CCARH collection. The authors acknowledged the limits of the closed set formulation, stating "the possibility that a composer not represented in the dataset wrote the piece should be kept open" \citep{vankranenburg2008measuring}.

Regarding the case of Josquin's works, the originating publication \citep{brinkman2016musical} did not perform careful musicological research to restrict the set of possible authors. Instead, reliance was placed on a musicologically curated dataset containing various well-known Renaissance composers collected by the Joquin Research Project (JRP)\footnote{\url{https://web.archive.org/web/20240724225841/https://josquin.stanford.edu/about/}}
. This approach enabled the automatic attribution of labels to more than 200 questionable works. While this method is more suitable for large-scale problems, it lacks the reliability of a meticulous musicological analysis that narrows the possible attribution labels. Observing the evaluation procedure, the originating work employed a single hold-out strategy, resulting in a balanced accuracy of 50\%. A subsequent study \citep{mckay2018jsymbolic} adopted a 10-fold cross-validation procedure and evaluated the $F_1$-measure, which partially accounts for dataset imbalance. This study achieved 91\%  and 82\% accuracy for two different two-classes problems (Josquin vs. Ockeghem and Josquin vs. La Rue respectively). Additionally, the average balanced accuracy of a 10-fold cross-validation performed on the six classes was computed by Verma and Thickstun, yielding a value of 77\% \citep{verma2019convolutional}. Overall, recent improvements allow for a more rigorous approach to the problem, but the attribution of suspicious Josquin works remains an open challenge. Even in this case, the authors rely on a weak process for data collection, with only one musicologist responsible for the transcription -- thus with no external review -- and no musicological variants encoded: from the information available in the JRP documentation, it is not clear how dubious passages have been transcribed and encoded and how potential biases from the transcriber (e.g., systematic interpretation or error patterns) have been mitigated. However, \cite{mckay2018jsymbolic} put a particular attention on the automatic handling of confounding factors deriving from encoding choices, which can partly mitigate the problem.

In the case of the disputed attribution of Lennon-McCartney song portions, the originating work by \cite{glickman2019data} approached the problem by analyzing symbolic representations. Specifically, they utilized published music score editions of 70 songs by The Beatles as their primary source material. From these symbolic scores, they manually extracted a set of hand-crafted features designed to capture compositional style, distinct from performance or recording artifacts. These features were then used to train a logistic linear regression model with both $L_1$ and $L_2$ regularization factors. The model's performance was evaluated using standard accuracy in a leave-one-out validation scheme, achieving 76\% correct predictions for disputed song excerpts. This study is notable as one of the few prominent computational authorship attribution investigations in popular music that operates on symbolic data, employing a methodology of feature extraction from scores comparable to approaches used for classical music. However, the result indicates that the    re-attribution of these song excerpts remains an open challenge. The authors, interested in establishing ``the  stylistic fingerprint of a songwriter based solely on a corpus of songs’ musical content'', based their training on existing published collections. The study did not delve into a detailed investigation of potential editorial choices within these music transcriptions, which can be a complex factor for popular music that often has a strong oral tradition alongside published forms. Future work could enhance the validation and feature extraction using more recent tools and methodologies.

\section{Takeaways for future research in composer identification}
\label{sec:guidelines}

This section summarizes the major findings of the research that may be useful to the MIR community for further investigations in the analysis of musical styles and especially for improving composer identification models.

\subsection{Number of classes}

While reducing the number of classes or merging stylistically similar ones can artificially boost accuracy, such modifications must be approached with caution, as they can alter the fundamental research question. For instance, in the study by \cite{brinkman2016musical}, changing class definitions would shift the inquiry from "\textit{who is the author of these works?}" to "\textit{is Josquin the author of these works?}". Ultimately, adherence to the robust evaluation protocols detailed in Section~\ref{sec:evaluation} is key to obtaining reliable numerical results whose susceptibility to variations in the number of classes is minimized.

\subsection{Historical period length}

The interpretation of dataset period length must consider the research goal. For some benchmarking purposes, a wide historical period might be intentionally chosen to test a model's ability to distinguish very disparate styles, potentially making the task easier if the number of classes is small and the stylistic differences are gross. However, for musicological authorship attribution, particularly when attempting to distinguish between contemporaneous composers or those from a similar school, a narrow historical focus is crucial. Using a wide historical span in an attribution context risks the model learning period-specific or broad genre-specific cues rather than the nuanced, individual stylistic traits of a composer. Thus, while an experiment using a wide period might be valid for its stated benchmarking goal, its resulting model may have limited utility or generalizability for fine-grained attribution tasks.

\subsection{Number of instances}

The dataset should be balanced in terms of the number of instances per class. Imbalances, especially in large datasets, can negatively impact the evaluation stage, particularly for less reliable classes. Moreover, the number of instances should be a trade-off between a large number for improving the generalization ability of the model and a small number to allow curated data collection, which is often costly and time-consuming. Developing models using not well curated data can generate inaccurate systems that are hardly usable in different datasets.

\subsection{Problem formulation}
In the analyzed literature, composer identification experiments were carried out among a finite set of composers. This is the "closed set" formulation of the authorship attribution problem as defined by \cite{juola2007authorship}. However, a more appropriate choice of model architecture and data analysis could pave the way for the "open set" formulation, where an additional class "other" is added. The use of a residual class allows researchers to check for the validity of the possible attributions included in the dataset. The construction of the residual class is challenging and is likely the reason why the literature has not adopted this approach. However, Bayesian machine learning offers a different perspective that is useful to the "open set" formulation.

In the typical frequentist formulation, the model predicts a point-estimate of the possible composer, while in the Bayesian formulation, the model predicts a probability distribution over the possible composers. This distribution can be used to compute the probability of the "other" class, which is the probability that the model is uncertain about the attribution. While an uncertainty measure can also be computed from frequentist models, the Bayesian approach offers the opportunity to disentangle aleatory and epistemic uncertainty \citep{depeweg2017decomposition,valdenegro-toro2022deeper}. Epistemic uncertainty is tied to the data seen during training, while aleatory probability is connected with some intrinsic error in the data, due to, for instance, the transcriber errors, the feature extraction, or the encoding system. In an authorship attribution case, having a large epistemic uncertainty for a certain musical work means that the work is not represented in the train set. This is a clear indication that the model is not able to predict the authorship of the work given the information encoded in the dataset. Thus, a large epistemic uncertainty means that the music style of the input music is different from any composer as represented in the dataset.

In the context of MIR and the categorization of large databases, the "open set" formulation could be particularly useful when the dataset is collected from different sources using automated scraping, consequently being subject to possible encoding incoherence and errors. In general, it allows testing the homogeneity of the train and test sets, thus allowing for further checks during the use of the model in real-world applications.

\subsection{Data Collection}
The data integrity has a fundamental role in the validity of the experiments. In the existing literature about authorship attribution cases, no author has paid the worth attention to the data collection protocol. The same happens in the more general context of composer identification when large datasets are collected for training models that are used on different data. Here, I reference \cite{pugin2015challenge} and \cite{sculley2008meaning} for a detailed description of a musicological approach to data collection, that can be summarized in the following recommendations:
\begin{itemize}
	\item Ensuring data quality and transparency by continuous validation and correction by multiple independent experts while making underlying assumptions explicit.
	\item Promoting interoperability and standardization by using widely accepted and extensible standards, such as MIDI, MEI, and MusicXML, to ensure compatibility across different research domains and datasets \citep{ludovico2019adoption}.
	\item Applying diverse methodologies and data representations to capture different aspects of complex problems, avoiding reliance on a single data model. This includes different file formats and multiple representations for distinguishing levels of editorial interventions.
	\item Facilitating open access and reproducibility by making data and methods available whenever possible, ensuring that research can be independently verified and built upon.
\end{itemize}

\subsection{Encoding formats}
It is also possible deriving some insights by examining the encoding formats discussed in the literature. According to the author's experience, musicologists emphasize the comprehensive representation of all musical elements, including accidentals, editorial interpretations, variations, dynamics, and more. Consequently, the musicological approach often harbors skepticism towards formats perceived as overly simplistic, such as MIDI. While this skepticism may have some theoretical basis, the absence of technical tools to fully utilize the detailed information that musicologists value remains a significant barrier \citep{simonetta2023optimizing,zhang2023symbolic}. Excessive effort in gathering highly detailed data may be a misallocation of resources that could be better spent on eliminating potential confounding factors from editorial and encoding choices from the data as well as enlarging the size of the dataset.

\subsection{Validation protocols}
Regarding the validation protocols, many studies used a simple hold-out scheme or inaccurate measures of merit.
While measures solely based on precision and recall may be meaningful, they do not offer a comprehensive representation of the reliability of the model.
Moreover, using a simple hold-out split for validating the model, when combined with the typical data scarcity problem connected with curated musicological corpuses, likely lead to imprecise evaluation measures.
As discussed in Section~\ref{sec:evaluation}, it is recommended to use $k$-fold validation with BA or MCC as measures of merit.

\subsection{Pondering claims}

In general, authors often overstate the effectiveness of their models, which can result in questionable attributions. For example, BWV 534 was initially credited to J.L. Krebs by \cite{vankranenburg2005musical}, but this attribution was later reconsidered and withdrawn by the same researchers \citep{vankranenburg2008measuring}.

A claim too much optimistic may influence the future research in the field, leading to a waste of resources and time. It is important to be cautious in the interpretation of the results and to always consider the limitations of the model.
In general, a musical work should be considered as re-attributed only after multiple meticulous studies using different data collections, classes, and tests.

It must be kept in mind that the accuracy measure is not the sole aspect of the evaluation, where data integrity, experimental protocol, and research question play a fundamental role.
It is the opinion of the authors that computational methods alone cannot suffice for authorship attribution problems. Their predictions must be carefully interpreted and guided by domain experts.

It is worth mentioning that the repertoire studied by the literature is extremely focused on the famous German composers (Bach, Mozart, Haydn, Beethoven). Little has been studied about other composers, thus limiting the understanding of the full potential of automated composer identification for authorship attribution problems.

\section{Conclusions}

This systematic review has meticulously surveyed the landscape of composer identification and authorship attribution from symbolic music scores, drawing critical insights from 58 peer-reviewed studies spanning decades of computational musicology. My analysis underscores a pivotal challenge: while technological advancements have introduced sophisticated machine learning models, the reliability and musicological validity of attribution claims often remain compromised by inconsistent evaluation practices and inherent dataset limitations.

Key findings highlight the critical importance of robust evaluation metrics, particularly Balanced Accuracy and MCC, over traditional accuracy, especially when dealing with imbalanced datasets common in this domain. I identify a pervasive reliance on simple hold-out validation and a concentration on a limited repertoire of Western classical composers, often overlooking crucial aspects of data integrity and musicological curation. The case studies on Bach, Josquin Desprez, and Lennon-McCartney illustrate the complex interplay between computational predictions and musicological interpretation, revealing that even high classification scores do not automatically translate to definitive authorship.

To address these gaps, this work proposes a comprehensive set of guidelines for future research. These recommendations emphasize the necessity of transparent, musicologically informed data collection protocols, the adoption of rigorous cross-validation strategies, and the judicious interpretation of model outputs, ideally within a Bayesian framework to quantify uncertainty. Ultimately, while computational methods offer powerful tools for stylistic analysis, their true potential for authorship attribution can only be realized through interdisciplinary collaboration and a cautious, evidence-based approach that integrates machine learning insights with deep musicological expertise.

\section*{Acknowledgments}
I am indebted to the anonymous reviewers for their exceptionally thorough and insightful critiques. Their guidance was instrumental in reshaping this work, and the manuscript has been substantially strengthened as a direct result of their generous contribution of time and expertise.

\section*{Funding Statement}
This work has been funded by the European Union (Horizon Programme for Research and Innovation 2021-2027, ERC Advanced Grant “The Italian Lauda: Disseminating Poetry and Concepts Through Melody (12th-16th century)”, acronym LAUDARE, project no. 101054750). The views and opinions expressed are, however, only those of the author and do not necessarily reflect those of the European Union or the European Research Council. Neither the European Union nor the awarding authority can be held responsible for such matters.
\begin{figure}[H]
    \centering
    \includegraphics[width=0.75\linewidth]{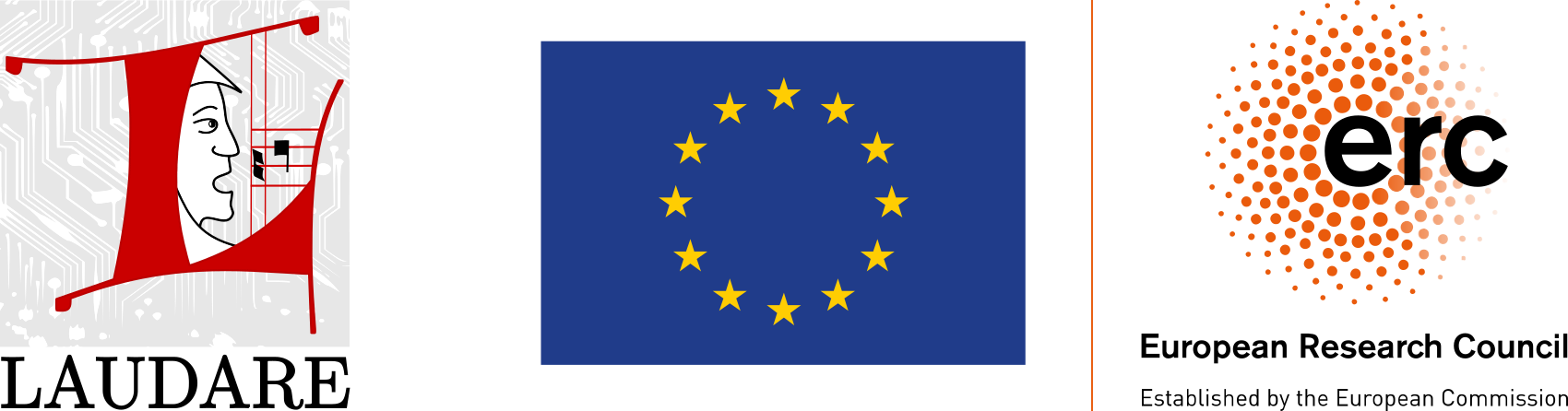}
\end{figure}

\bibliography{bibliography,zotero}

\end{document}